\documentclass[useAMS]{mn2e}

\usepackage{graphicx}	
\usepackage{amsmath}	
\usepackage{amssymb}	
\usepackage{multicol}       
\usepackage{bm}		
\usepackage{pdflscape}	
\usepackage[utf8]{inputenc}
\usepackage[T1]{fontenc}
\usepackage{ae,aecompl}

\title[A Tully-Fisher relation for GCs] {Asymptotic kinematics of Globular Clusters:
the emergence of a Tully-Fisher relation.} 

\author[X. Hernandez, A. J. Lara-D.I.] {X. Hernandez, A. J. Lara-D.I.\\ 
Instituto de Astronom\'{\i}a, Universidad Nacional Aut\'{o}noma de M\'{e}xico,
  Apartado Postal 70--264 C.P. 04510 M\'exico D.F. M\'exico. \\
}

\date{Released 11/09/2019}

\begin{document}

\label{firstpage}

\maketitle

\begin{abstract}
  Using a recent homogeneous sample of 40 high quality velocity dispersion profiles for Galactic globular clusters, we study
  the low gravitational acceleration regime relevant to the outskirts of these systems. We find that a simple empirical profile
  having a central Gaussian component and a constant large radius asymptote, $\sigma_{\infty}$, accurately describes the variety
  of observed velocity dispersion profiles. We use published population synthesis models,
  carefully tailored to each individual cluster, to estimate mass to light ratios from which total stellar masses, $M$, are
  inferred. We obtain a clear scaling, reminiscent of the galactic Tully-Fisher relation of $\sigma_{\infty}( km s^{-1})=
  0.084^{+0.075}_{-0.040} (M/M_{\odot})^{0.3 \pm 0.051} $, which is interesting to compare to the deep MOND limit of $\sigma_{\infty}
  (km s^{-1})=0.2(M/M_{\odot})^{0.25}$. Under a Newtonian interpretation, our results constitute a further restriction on models
  where initial conditions are crafted to yield the outer flattening observed today. Within a modified gravity scheme, as the
  globular clusters studied are not isolated objects in the deep MOND regime, the results obtained point towards a modified gravity
  where the external field effect of MOND does not appear, or is much suppressed.

\end{abstract}


\section{Introduction} \label{intro}

A topic of current debate is whether the well established gravitational anomalies evident in the rotation curves of
spiral galaxies, are due to the presence of dominant halos of as yet undetected dark matter, or perhaps indicative
of a change in gravitational physics appearing in the low acceleration regime. It has been shown that these anomalies
become conspicuous at acceleration scales below $a_{0}=1.2 \times 10^{-10} m s^{-1}$ (e.g. Famaey \& McGaugh 2012,
Lelli et al. 2017), where $a_{0}$ is the critical acceleration of MOND, Milgrom (1983). The central feature of MOND and
related modified theories of gravity constructed so as not to require the hypothesis of dark matter, are for centrifugal
equilibrium velocities about a baryonic mass $M$, which become flat at a value of $V_{TF}=(G M a_{0})^{1/4}$ for large radii.
The corresponding expectation for the velocity dispersion velocities of isothermal pressure supported systems is
$\sigma_{TF}=3^{-1/2}(G M a_{0})^{1/4}$, $\sigma_{\infty}( km s^{-1})= 0.2 (M/M_{\odot})^{0.25}$ e.g. McGaugh \& Wolf(2010).

Globular clusters (henceforth GCs) offer an interesting independent test of the generality of the situation encountered in rotation curves
of spiral galaxies, as they are pressure supported systems where under standard cosmology no significant dark mater presence
is expected, and towards their outskirts, reach the low acceleration regime of $a<a_{0}$. Starting with the work of Scarpa et al.
(2003) and Scarpa et al. (2007), is has become apparent that beyond the $a<a_{0}$ threshold, the observed projected velocity dispersion
profiles of Galactic globular clusters, are consistent with tending towards a finite value, as has been confirmed by e.g.
the analysis of the Lane et al. (2009), Lane et al. (2010) and Lane et al. (2011) data of Hernandez et al. (2013), and Sollima et al.
(2016), $\sigma_{R \rightarrow \infty}=\sigma_{\infty}$.

This has been interpreted as
evidence in favour of a modified gravity regime appearing in the $a<a_{0}$ regime by e.g. Hernandez et al. (2013) and Hernandez
et al. (2017), but explained under a Newtonian scenario through selecting initial conditions in terms of density profiles, binary
fractions and distributions and initial stellar mass functions for dynamical models, which evolve into the situation observed today.
Examples of the latter case include Claydon et al. (2017) who consider the contribution of unbound stars to the resulting present day
velocity dispersion profiles, obtaining an outer flattening consistent with observations, as was also found by Kennedy (2014) by
considering chaotic internal dynamics resulting from the interaction with the overall galactic potential, an effect which naturally
increases with decreasing GC mass.

As an independent line of enquiry, one can go from requiring a flattening of the present day velocity dispersion profile,
to exploring the mass scaling of $\sigma_{\infty}$, and compare to the expectations
under Newtonian or modified gravity. In Hernandez et al. (2013) some of us used available HST velocity dispersion profiles
for 8 GCs, and found a $\sigma_{\infty}\propto M^{0.31 \pm 0.06}$ scaling, for masses inferred from stellar population synthesis
models by McLaughlin \& van der Marel (2005), not including any dynamical assumptions.

In this paper we take advantage of the availability
of the GC velocity dispersion profile catalogue of Baumgardt et al. (2019), which includes uniformly reduced
high quality ground based ESO and Keck spectra from Baumgardt \& Hilker (2018), as well as Gaia DR2 results, to construct
the most recent and complete GC velocity dispersion profile library available. We also use stellar population modelling
carefully tailored to the specific metalicities and ages of each GC treated from McLaughlin \& van der Marel (2005) to
obtain mass to light ratios and masses, to determine the empirical scaling of $\sigma_{\infty}$ and $M$. For a sample
of 40 GCs with velocity dispersion profiles having a large radial coverage allowing accurate $\sigma_{\infty}$ inferences,
we obtain  $\sigma_{\infty}( km s^{-1})= 0.084^{+0.075}_{-0.040} (M/M_{\odot})^{0.3 \pm 0.051}$.

This paper is organised as follows: section 2 details the sample selection and velocity dispersion profile fitting,
where we also compare the fitted profiles to the observed velocity dispersion observations. In section 3 we discuss the resulting
$\sigma_{\infty}$ vs. $M$ scalings obtained, showing an equivalent globular cluster Tully-Fisher relation, and section
4 states our conclusions.

\section{Sample selection and velocity dispersion profile modelling}

That it only recently became apparent that classical Newtonian King models are insufficient to describe extended velocity dispersion
profiles of GC, is equivalent to the fact that the readily accessible to observation, central velocity dispersion values
of these systems, are perfectly consistent with Newtonian expectations. Indeed, observed GC velocity dispersion profiles are
characterised by a relatively constant central region, followed by a gradual drop, in consistency with expectations for isolated,
equilibrium Newtonian systems, which however, are followed by a convergence to a finite asymptotic value. These profiles can be
accurately described by the empirical profile:


\begin{figure*}
\vskip -37pt
\hskip -10pt \includegraphics[height=17cm,width=18.5cm]{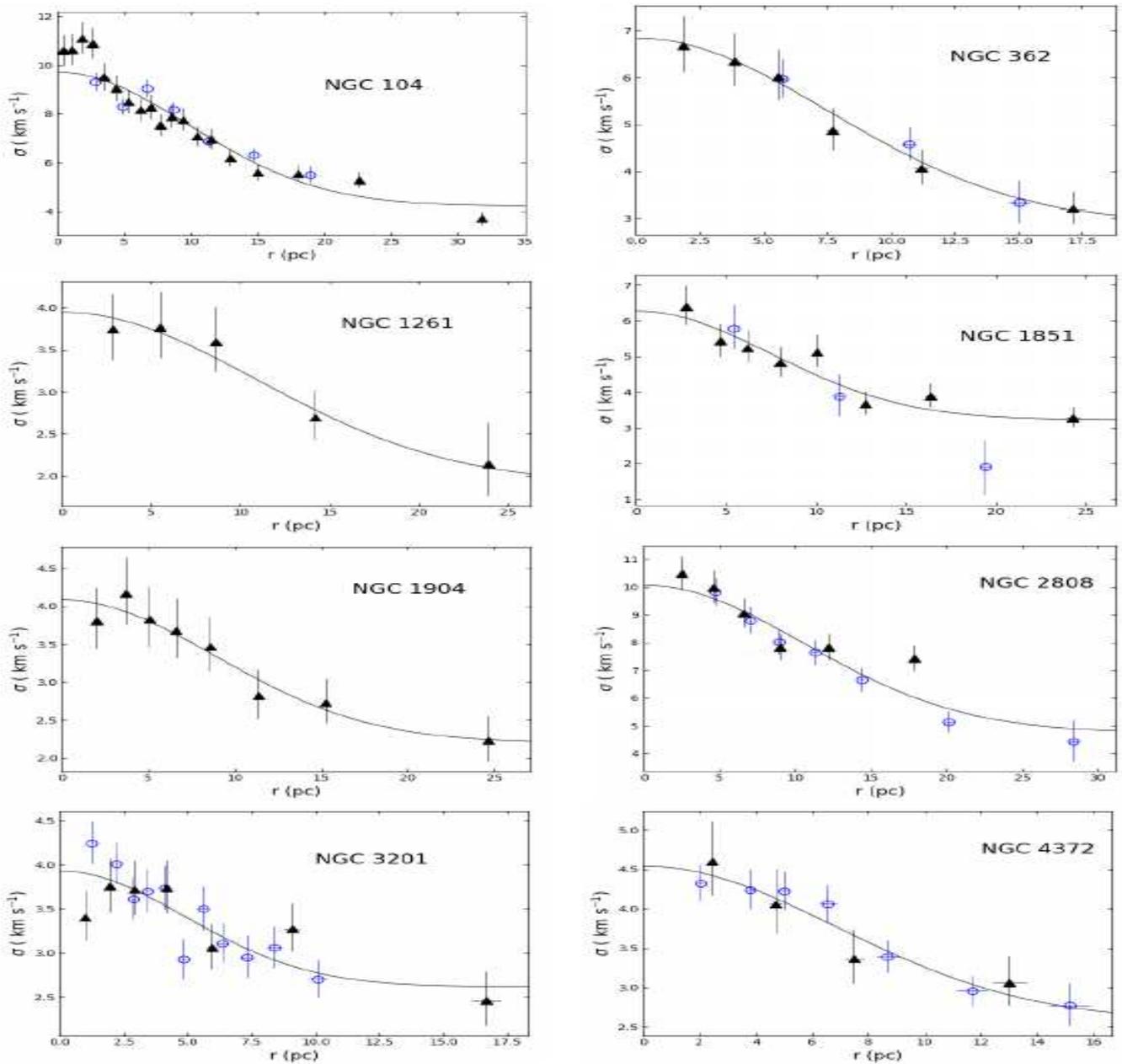}
\caption{The figure shows the first eight velocity dispersion profile fits to equation (1). The solid triangles show
ESO Keck radial dispersion velocity measurements, and the circles give Gaia DR2 data, with no systematic offset being present
between the two, from the Baumgardt et al. (2019) catalogue. That a good representation of the empirical profiles is afforded by
equation (1) is evident.}
\end{figure*}

\begin{figure*}
\vskip -37pt
\hskip -10pt \includegraphics[height=17cm,width=18.5cm]{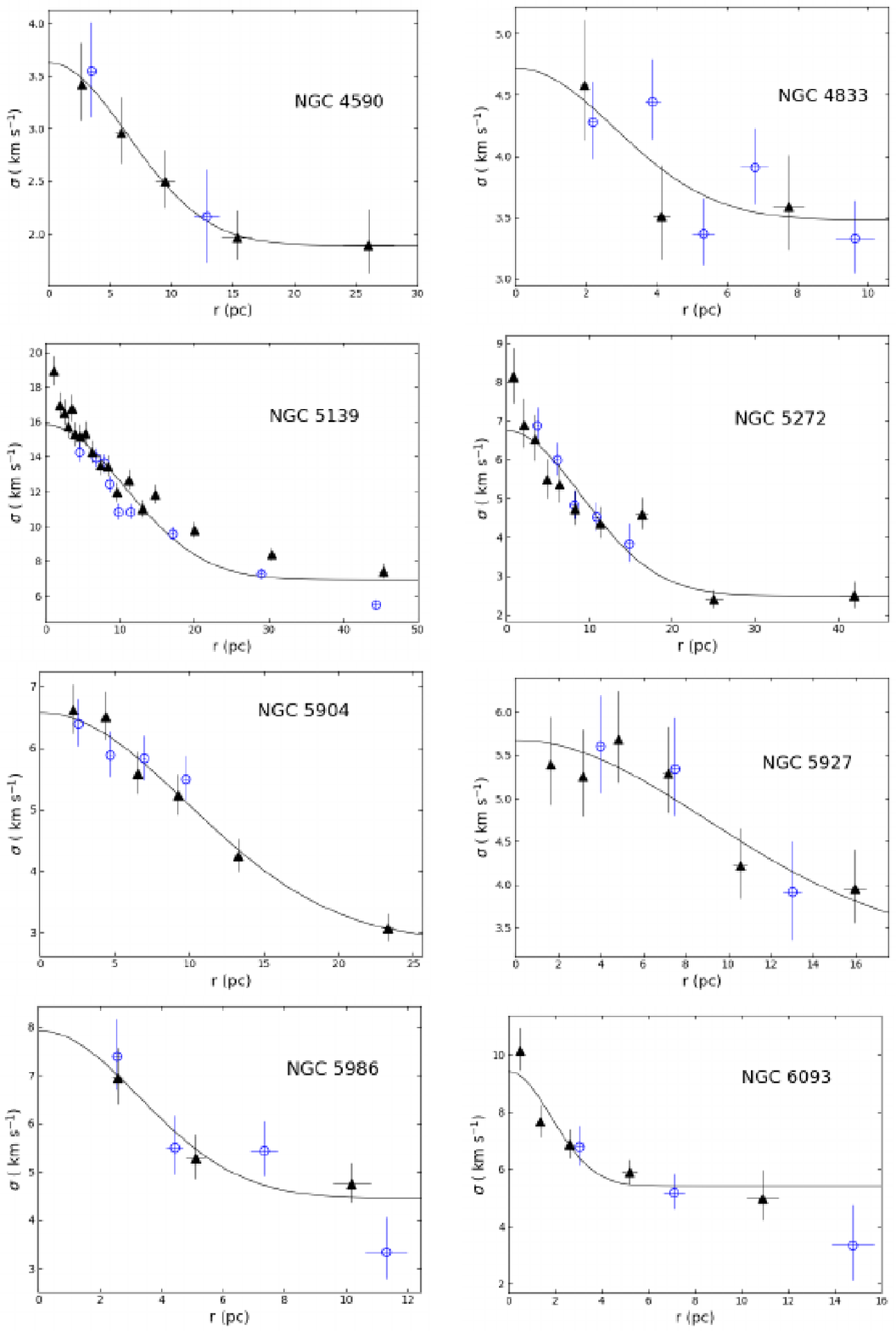}
\caption{The figure shows the second eight velocity dispersion profile fits to equation (1). The solid triangles show
ESO Keck radial dispersion velocity measurements, and the circles give Gaia DR2 data, with no systematic offset being present
between the two, from the Baumgardt et al. (2019) catalogue. That a good representation of the empirical profiles is afforded by
equation (1) is evident.}
\end{figure*}

\begin{figure*}
\vskip -37pt
\hskip -10pt \includegraphics[height=17cm,width=18.5cm]{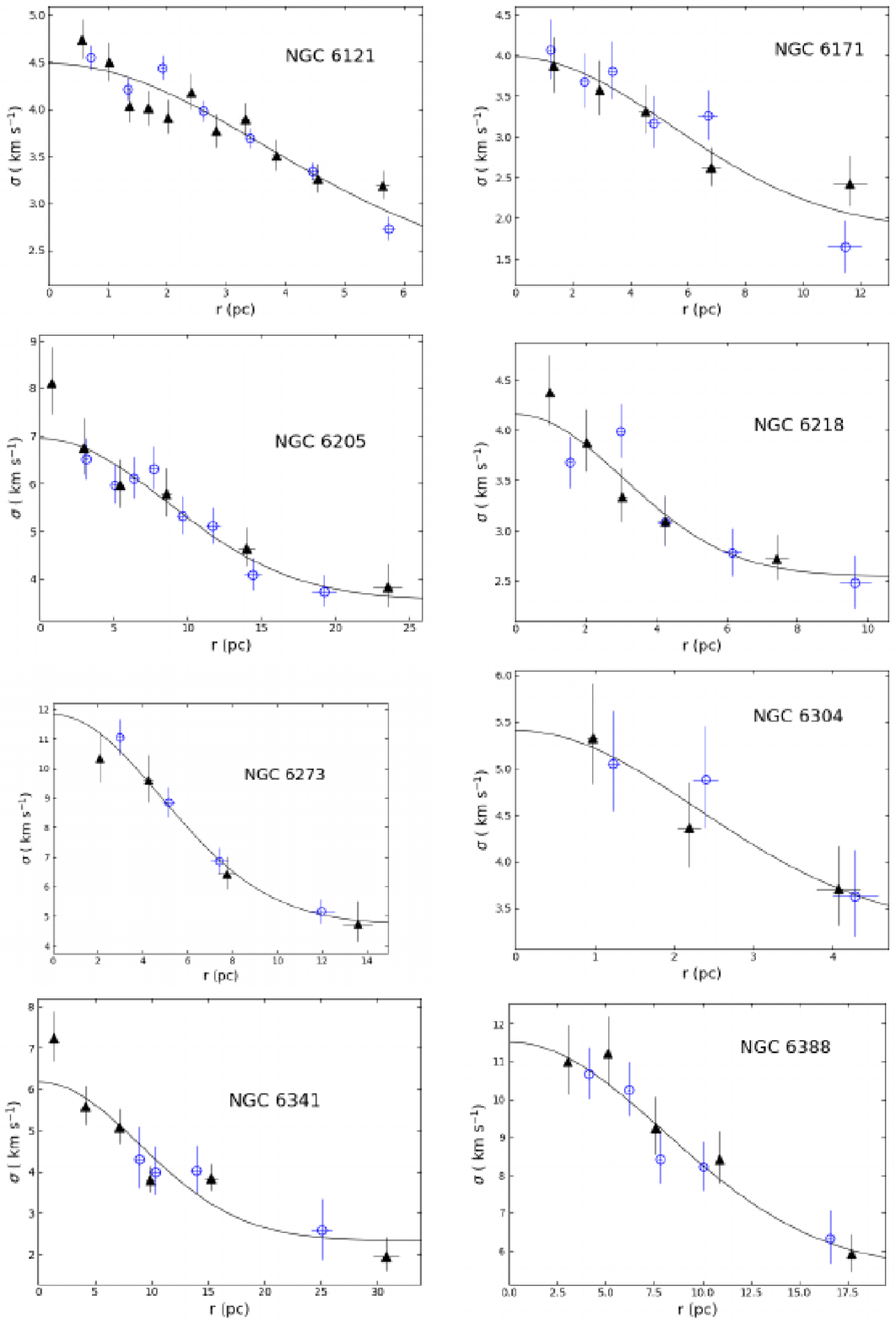}
\caption{The figure shows the third eight velocity dispersion profile fits to equation (1). The solid triangles show
ESO Keck radial dispersion velocity measurements, and the circles give Gaia DR2 data, with no systematic offset being present
between the two, from the Baumgardt et al. (2019) catalogue. That a good representation of the empirical profiles is afforded by
equation (1) is evident.}
\end{figure*}

\begin{figure*}
\vskip -37pt
\hskip -10pt \includegraphics[height=17cm,width=18.5cm]{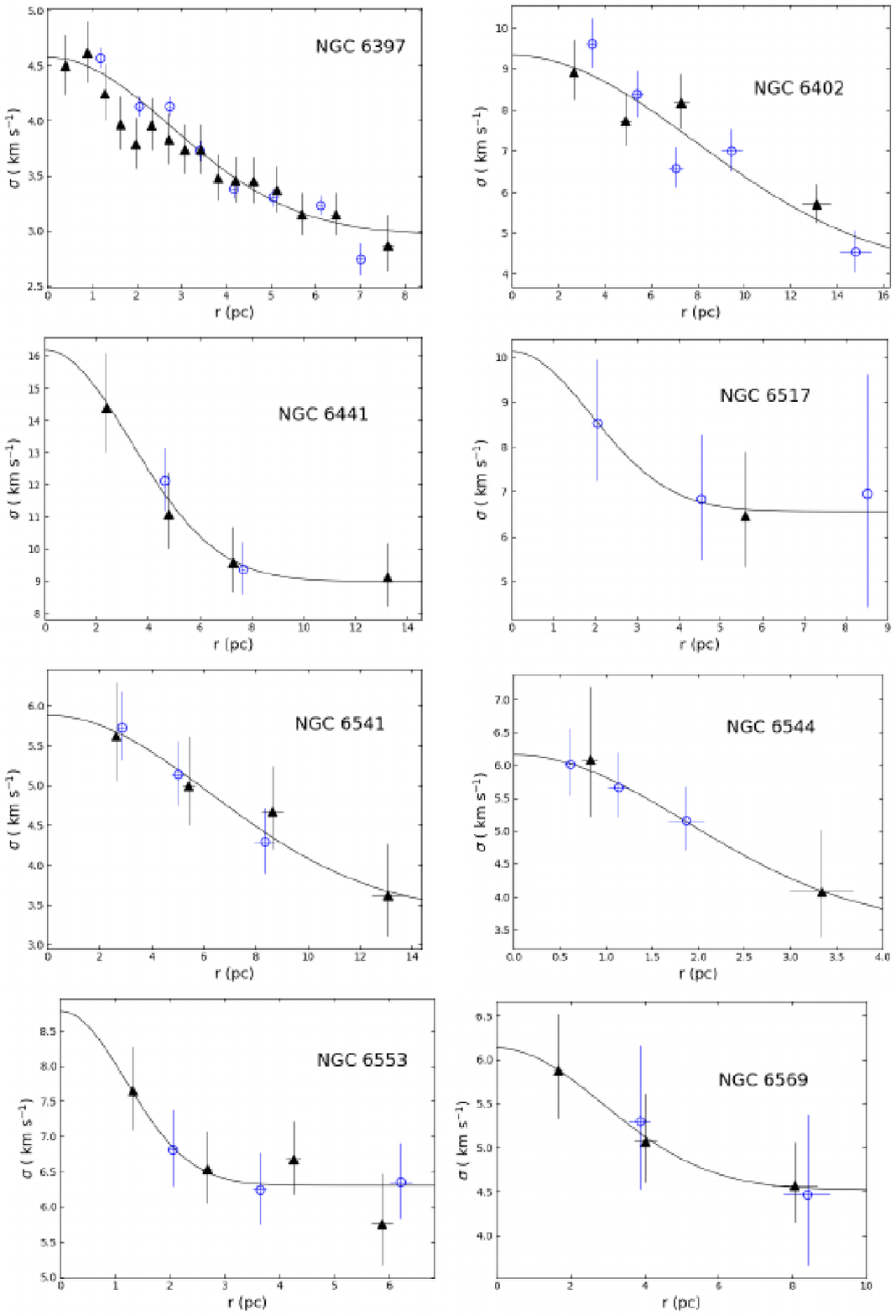}
\caption{The figure shows the fourth eight velocity dispersion profile fits to equation (1). The solid triangles show
ESO Keck radial dispersion velocity measurements, and the circles give Gaia DR2 data, with no systematic offset being present
between the two, from the Baumgardt et al. (2019) catalogue. That a good representation of the empirical profiles is afforded by
equation (1) is evident.}
\end{figure*}

\begin{figure*}
\vskip -37pt
\hskip -10pt \includegraphics[height=17cm,width=18.5cm]{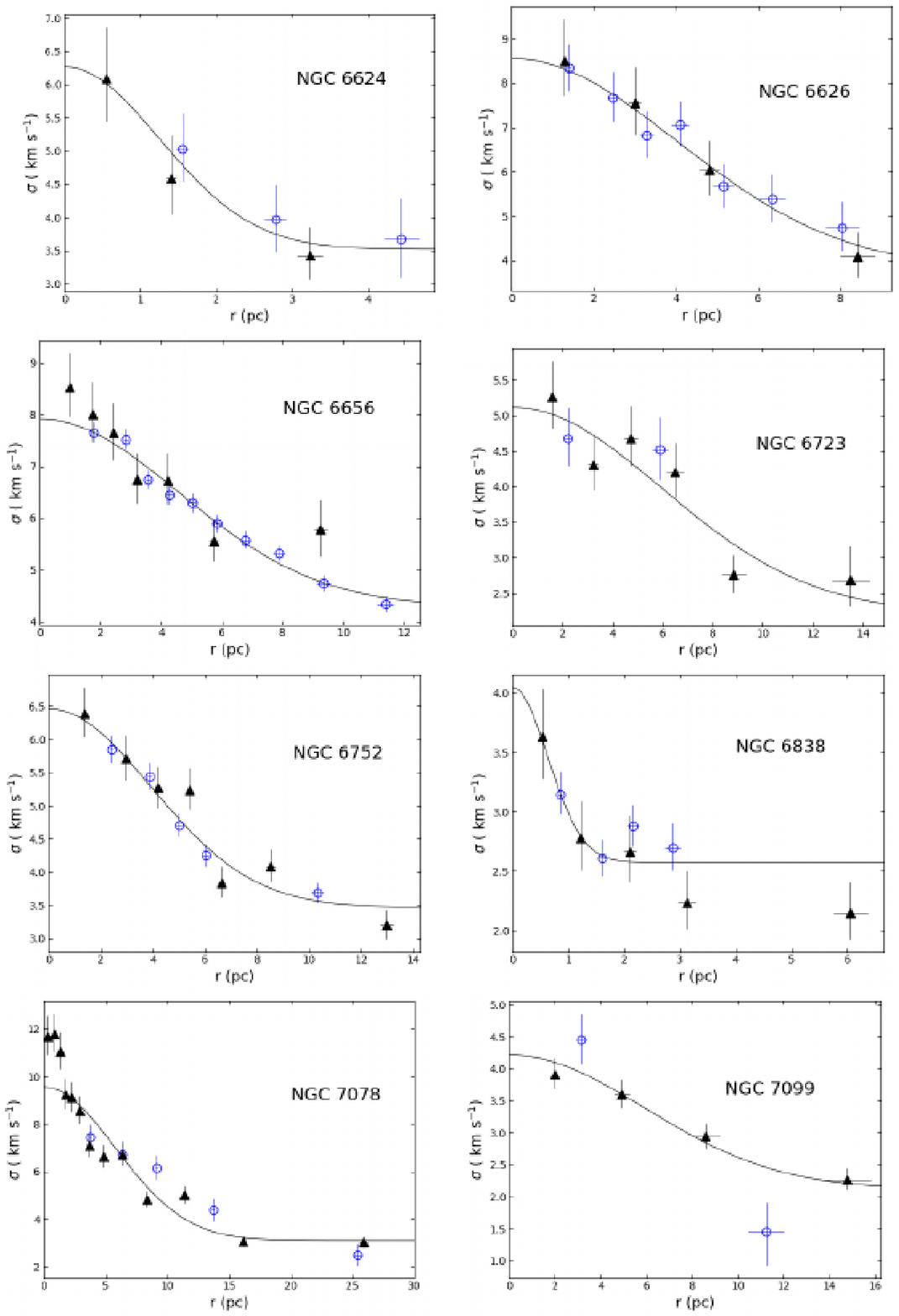}
\caption{The figure shows the fifth eight velocity dispersion profile fits to equation (1). The solid triangles show
ESO Keck radial dispersion velocity measurements, and the circles give Gaia DR2 data, with no systematic offset being present
between the two, from the Baumgardt et al. (2019) catalogue. That a good representation of the empirical profiles is afforded by
equation (1) is evident.}
\end{figure*}

\begin{equation}
\sigma(R)=\sigma_{1}e^{-(R/R_{\sigma})^{2}} +\sigma_{\infty}
\end{equation}

The above profile was introduced in Hernandez et al. (2012), where is was shown to provide an accurate description to the empirical
velocity dispersion profiles of a small sample of 8 Galactic globular clusters. Indeed, in Durazo et al. (2017) and Durazo et al.
(2018), the same empirical profile was applied to pressure supported systems at scales orders of magnitude larger than those of GCs;
the velocity dispersion profiles of over 300 low rotation elliptical galaxies from the CALIFA and MANGA samples, carefully traced by
the proposed empirical equation.  It is through fits to equation (1) that we will characterise the velocity dispersion profiles of
the GCs sample used.

As there are 3 free parameters in equation (1), we need at least three radial points in any profile to be fitted. So as to guarantee
accurate fits, we consider only GCs with profiles including at least four radial points. This cut reduces the Baumgardt et al.(2019)
catalogue to 69 of the 139 GCs  reported. Still, large error bars in profiles having only 4, 5 or 6 points, yield fits with very
large confidence intervals on the three parameters sought. In the interest of constructing a high quality sample, we exclude  
profiles where the relative error on any of the fitted parameters is larger than 0.6. This high quality cut removes a further 29 GCs,
leaving us with our final sample of 40 objects.


After taking distances to the systems studied consistently from Baumardt et al. (2019), { we implement a Chi squared fitting
procedure to obtain best fit parameters to equation (1), together with their corresponding confidence intervals. Figures (1)
to (5) show the fitted profiles for our whole sample of 40 GCs. The sample shows a diversity of profiles covering a range of central
velocity dispersion values from the less than $4 km s^{-1}$ of NGC 4590, to the more than $16 km s^{-1}$ of NGC 6441, a range of 
asymptotic velocity dispersion values, from the  $2 km s^{-1}$ of NGC 6171 to the more than  $7 km s^{-1}$ of NGC 5139, as well
as a varied morphology in terms of quotients between the central and asymptotic velocity dispersion values and numbers of points
in each profile.}

As can be seen from the figures, the
empirical profile being fitted provides an accurate description across the various parameters of the sample, with the small caveat of
a slight central overshooting appearing in a few of the high central velocity dispersion cases. This last however, does not
in any way affect the accuracy with which $\sigma_{\infty}$ is inferred, which is the parameter we are interested in here. Solid
triangles and empty circles show observations from ESO Keck radial velocity spectra and Gaia DR2 data, respectively. As there is
no systematic difference between the above two, we treat both data samples indistinctly throughout as comprising single { 1D
$\sigma(R)$ } profiles for each GC. 

Finally, to estimate the total stellar masses of the GCs in question, we turn to the detailed population synthesis models of
McLaughlin \& van der Marel (2005), where independent metallicity estimates and careful HR diagram comparisons are used to infer
the mass to light ratios in the V band for a very complete selection of Galactic GCs, fortunately including all the ones treated
here. In the above study the main uncertainty in the resulting mass to light ratios comes from the assumed stellar mass functions.
To account for this, we take as confidence intervals on the mass to light ratios of the GCs treated, the extremes of the three models
provided (excluding Salpeter mass functions, see below), and as central values, the means of the central values of the three models
provided. Notice that we do not use any dynamical models to determine total masses, so that the final parameters obtained are purely
empirical and fully independent of any assumptions on the structure of gravity at any scale. The total magnitudes in the V band we take
from the most recent update of the Harris catalogue, Harris (1996), 2010 edition, and distances consistently from Baumgardt et al. (2019)
to infer the total stellar masses for the GCs in our final sample, given in the last column of table 1.

\section{A Tully-Fisher relation for GCs}

\begin{figure}
\vskip -37pt
\hskip -10pt \includegraphics[height=7.0cm,width=8.5cm]{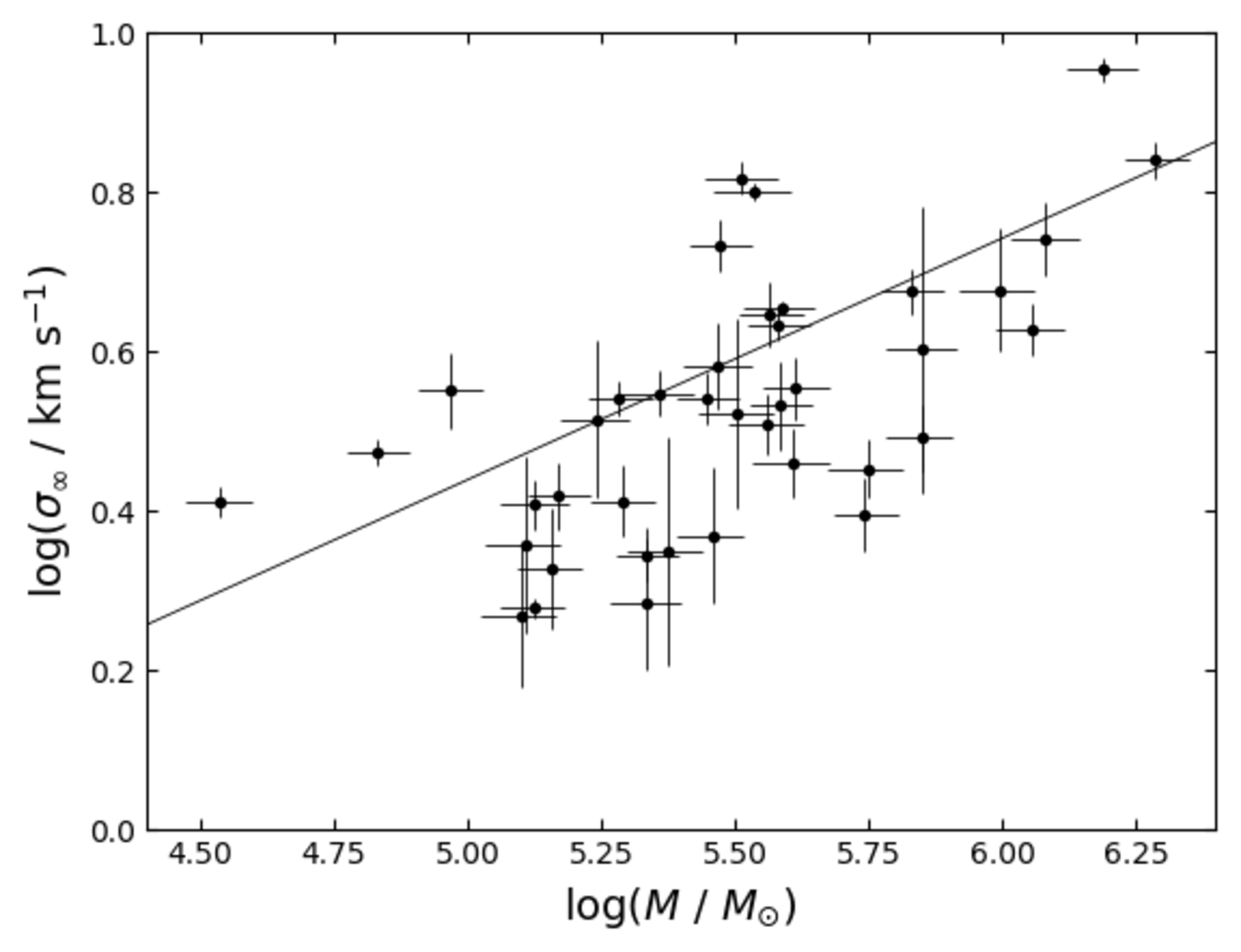}
\caption{The figure shows asymptotic large radii projected velocity dispersion values, $\sigma_{\infty}$, and corresponding
stellar population total mass estimates, $M$, for 40 Galactic globular clusters with high quality kinematic profile observations, on a
logarithmic plot. The line gives a best linear fit to the data, yielding  $\sigma_{\infty}( km s^{-1})= 0.084^{+0.075}_{-0.040}
(M/M_{\odot})^{0.3 \pm 0.051}$, which is consistent with deep MOND expectations of $\sigma_{\infty} (km s^{-1})=0.2(M/M_{\odot})^{0.25}$.}
\end{figure}

Having obtained the best currently available estimates for $\sigma_{\infty}$ and $M$ for the GCs in our sample, in
Figure (6) we give our final result for the 40 GCs treated, with corresponding error bars. The solid line in this figure shows a
linear best fit of  $\sigma_{\infty}( km s^{-1})= 0.084^{+0.075}_{-0.040} (M/M_{\odot})^{0.3 \pm 0.051}$. This is consistent with the expected
scaling of $\sigma_{\infty} (km s^{-1})=0.2(M/M_{\odot})^{0.25}$ for isolated pressure supported systems in MOND. Thus, the trend observed
in Figure (6) represents an equivalent empirical relation to the well known baryonic Tully-Fisher for disk galaxies, both consistent
with a fixed underlying trend, once the difference between full centrifugal support and close to isothermal pressure support is taken
into account. { We quantify the degree of correlation found in the data shown in Figure (2) through the value of the corresponding
Pearson's correlation coefficient, $r$, in this case of $r=0.64 (0.41, 0.80)$, where the numbers in brackets give the 95\% confidence
interval on the quoted value, a convention which we will use throughout. In spite of the presence of a correlation with parameters
consistent with MONDian expectations,} one difference between our result and the galactic Tully-Fisher relation is evident, the presence
of a significant intrinsic scatter in the data, well above what appears in spiral galaxies. 

{ We can now try to asses the possible presence of an external field effect, EFE, of standard versions of MOND, where the presence of an
external acceleration larger than the internal acceleration of an $a<a_{0}$ system essentially reverts the system in question to standard
Newtonian dynamics. This is suggested by the presence of a weak and noisy anti-correlation between $\sigma_{\infty}$ and the Galactocentric
radii of the clusters studied, $R_{G}$, also taken consistently from Baumgardt et al (2019). This correlation has a low Pearson's
$r=-0.41 (-0.64, -0.11)$, but however, the relevance of an EFE could be blurred by a spread in total mass, which is also present in the
systems studied. Thus, we now plot again Figure (6), but only for a sub-sample of GCs, those with the largest galactocentric distances,
15 clusters with $R_{G}> 7 kpc$. This is shown in Figure (7). We see that in spite of having a significantly reduced sample, the trend
remains consistent with MONDian expectations with a slope of $0.35 \pm 0.061$ (this time at a 1.46 sigma level), and the correlation
coefficient actually increases from what was obtained in Figure (6), this time yielding $r=0.8 (0.48, 0.93)$.

We could hence be seeing
the presence of an EFE to some level, not totally erasing the isolated MONDian
expectations which are consistent with the plot of the full sample, but which
could explain the appearance of a cleaner trend for the large RG clusters than
for the low RG ones.

Alternatively, the slight average increase in $\sigma_{\infty}$ with decreasing $R_{G}$ could be signaling the presence of disturbed
kinematics towards the outer regions of the GCs studied due to the effects of Galactic tides, naturally becoming more relevant
in going to smaller Galactocentric radii. We can explore this option by considering the tidal radii of the clusters in our sample,
$R_{T}$, also available in Baumgardt et al. (2019), carefully calculated within a Newtonian framework, and accounting for the detailed
orbits of each of the GCs reported, as constrained by Gaia proper motions. Since $\sigma_{\infty}$ is sensitive to the velocity
dispersion measurements available in a range of distances from the centres of the GCs treated of $R>(1-3)R_{\sigma}$, and as 
the effects of Galactic tides become more relevant as the ratio $(R_{\sigma}/R_{T})$ increases, if Galactic tides are strongly influencing
our $\sigma_{\infty}$ estimates, we would expect a strong positive correlation between these last two quantities. We now plot in Figure (8)
inferred values of $\sigma_{\infty}$ as a function of $(R_{\sigma}/R_{T})$ for our full sample. The result is a very noisy
{\it anti}-correlation, with a low significance implied by a correlation coefficient of $r=-0.38 (-0.62, -0.08)$. Thus we can dismiss a
strong contribution from Galactic tidal effects to the $\sigma_{\infty}$ values obtained.

We note also the presence of a positive correlation between $\sigma_{1}$
and $\sigma_{\infty}$. A power law fit to shows a best fit
Chi squared regression of $\sigma_{\infty} \propto \sigma_{1}^{(0.53 \pm 0.09)}$,
thus, on average, $\sigma_{\infty}$ values scale with the square root of 
$\sigma_{1}$ ones. This is quite interesting, as at approximately constant
half-light radii (or some other equivalent scale radii), the central
Newtonian regions of the GCs would be expected to show a scaling of
$\sigma(R=0) \propto M^{1/2}$. Under MONDian expectations, the outskirts of
the systems would show a TF $\sigma_{\infty} \propto M^{1/4}$ scaling, and hence,
central values of $\sigma$ (which are dominated by the $\sigma_{1}$ values)
which scale with the square of $\sigma_{\infty}$. The correlation coefficient
for this plot is of r=0.59 (0.34, 0.76), a lower value than what we found
for the TF plots, which could signal the blurring of the trend between
$\sigma(R=0)$ and $\sigma_{\infty}$ in going to the $\sigma_{1}$ vs. $\sigma_{\infty}$ space,
the presence of a range of actual characteristic values for the radius
relevant to the Newtonian virial determinations, or indeed the absence
of any actual physical correlation.


}

A number of caveats must be mentioned, firstly, it is now known that many Galactic globular clusters, if not all, show in the details some
evidence for the presence of multiple stellar populations (see e.g { the authoritative references Piotto et al. 2015 and Milone et al. 2017}
or Cordoni et al. 2019 and Pasquato \& Milone 2019 for two recent examples), so that the single stellar population assumption taken in
McLaughlin \& van der Marel (2005) to derive the mass to light ratios used, is only a first order approximation. Thus, the confidence
intervals on the mass estimates we use here are probably an underestimation. The same happens when considering the possibility of Salpeter
stellar mass functions, which do not alter the resulting power law fit, but do shift the data to the right by a small factor. Accounting for
the two above effects, to some varying degree in the different GCs, might also reduce the effective scatter observed and make the distribution
more compatible with a single power law.

\begin{figure}
\hskip -10pt \includegraphics[height=7cm,width=8.5cm]{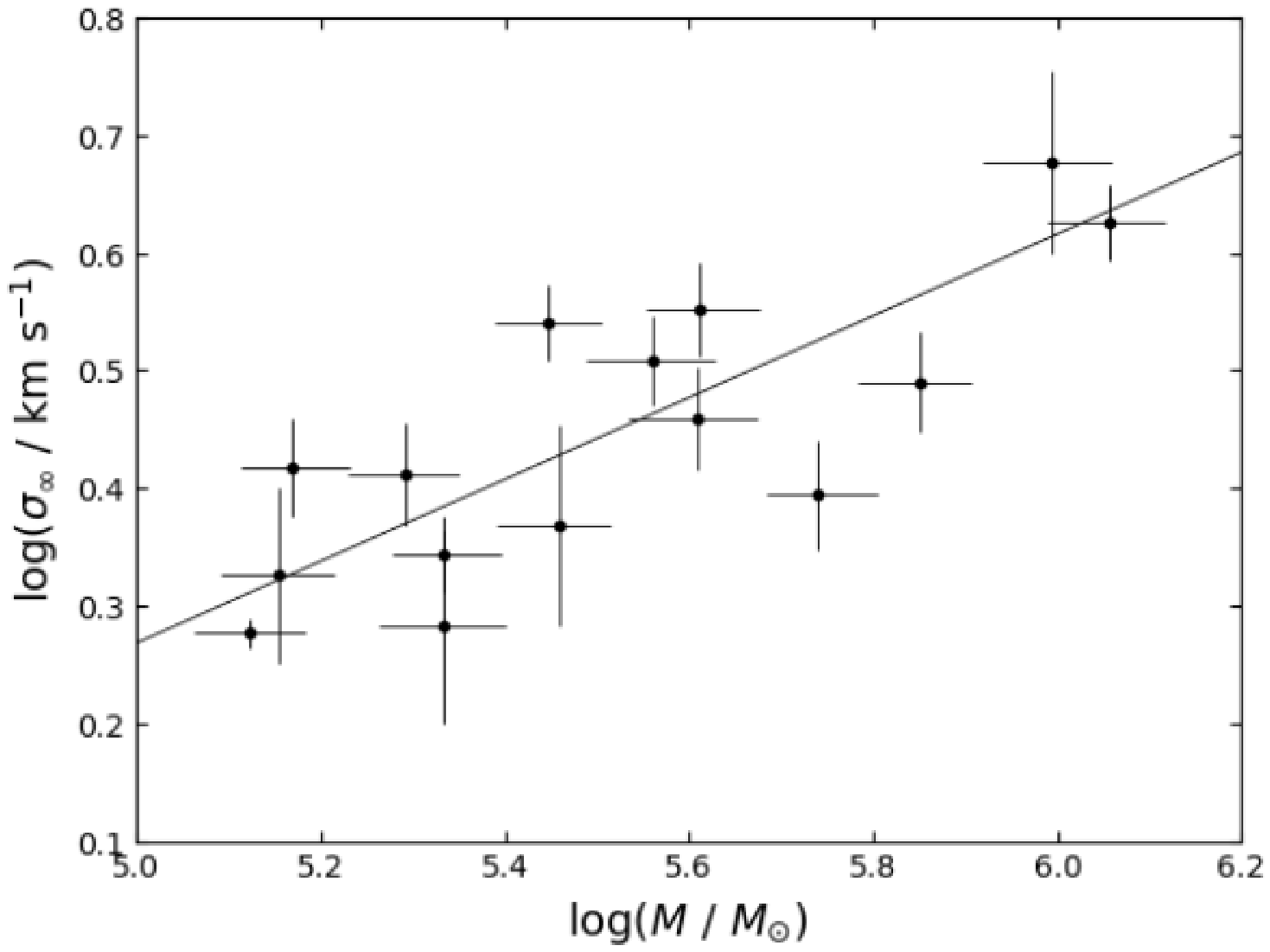}
\caption{The figure shows asymptotic large radii projected velocity dispersion values, $\sigma_{\infty}$, and corresponding
stellar population total mass estimates, $M$, for the 15 Galactic globular clusters with Galactocentric radii $>7kpc$, on a
logarithmic plot. The line gives a best linear fit to the data, yielding  $\sigma_{\infty}( km s^{-1})= 0.034^{+0.04}_{-0.02}
(M/M_{\odot})^{0.35 \pm 0.06}$, which is comparable to deep MOND expectations of $\sigma_{\infty} (km s^{-1})=0.2(M/M_{\odot})^{0.25}$.}
\end{figure}

{ Also, notice that in going towards the outer regions which crucially determine the $\sigma_{\infty}$ inferences, the velocity dispersion
profiles used are naturally determined from observations of stars tending towards the brighter end of the distribution found in the observed
systems, typically turn-off and evolved stars. However, due to mass segregation effects, the majority of stars in these regions are in fact
low-mass ones. This introduces a systematic under-estimate of the true $\sigma_{\infty}$ values, since due to partial energy
equipartition (and stellar evolution effects), these numerically dominant low-mass stars will have an enhanced velocity dispersion with
respect to that of the more massive stars on which the empirical estimate is based. It is interesting to note that modelling this effect
will necessarily result in an upwards shift of our empirical points in Figure (2), perhaps bringing the normalisation of
$0.084 ^{+0.075}_{-0.04}$ into a better agreement with the MONDian expectation of $0.2$.}

Also, the analogy with centrifugally supported disks breaks down when considering that in that case, the kinematically observed quantity
offers a very direct determination of the dynamics, beyond minor details introduced by effects such as asymmetric drift and gas pressure,
when considering radio observations. In the case of globular clusters however, stellar orbits are much closer to isothermal
distributions, where the stars observed at a given radius actually sample during their orbits a very large range of radii, and hence of
radial forces. The details of the above will be strongly dependent on the particular features of each cluster, such as orbital anisotropy
parameters (even radial variations in this quantity) and the degree of central concentration in the present day density profiles.
Thus, it is perfectly possible that in Figure (6) we are seeing the superposition of a series of 'Tully-Fisher' relations,
the disentangling of which will await more detailed data than the ones presently available.

It is however encouraging, that inspite of the presence of the complications mentioned above, a very clear trend appears, and it is
extremely interesting that this trend should be consistent, to within internal confidence intervals, with MOND expectations
for isolated systems. 

This result is a challenge for some of the Newtonian explanations for the observed outer flattening of the velocity dispersion profile of
globular clusters, such as the appearance of chaotic dynamics investigated by Claydon et al. (2017), or the influence of tidal fields
explored in Lane et al. (2012), both of which depend on the mass of the cluster such that the extra dynamical heating effects considered
diminish in importance as the mass of the cluster in question increases. Thus, being problematic the understanding of a clear
positive trend as the one shown in Figure (6), consistent with a Tully-Fisher scaling.

From the point of view of MOND as such, the situation
is not much different, as most of the clusters treated lie at relatively small galactocentric distances, and would hence be expected
to lie within the region where the external field effect dominates, and no significant departure from Newtonian dynamics would
be expected (e.g. Famaey \& McGaugh 2012). From this latter point of view, our results would point towards a modified MONDian
gravity theory where the external field effect is much less relevant (if at all) than in the case of MOND, e.g. Milgrom (2011).

{ Indeed, we do not intend to single out
or exclude any particular explanation, but merely to point out an
interesting feature of the data, highly reminiscent of the
$V \propto M^{1/4}$ scaling found at galactic level, and which in our opinion
merits further investigation.}

\begin{figure}
\vskip -37pt
\hskip -10pt \includegraphics[height=7cm,width=8.5cm]{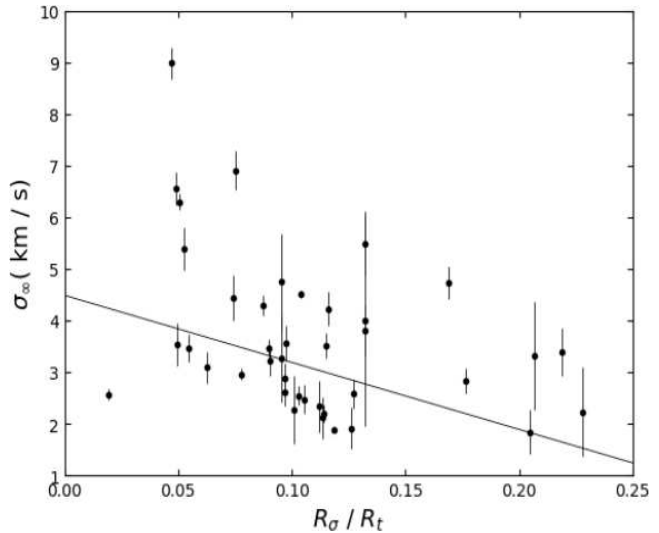}
\caption{Large radii asymptotic velocity dispersion inferences as a function of the ratio of $R_{\sigma}$ to the Newtonian tidal radii
of the GCs in our sample. The lack of any clear positive correlation, indeed, the presence of a very noisy negative trend, shown
by the solid line, indicates that our $\sigma_{\infty}$ inferences are unlikely to be significantly affected by dynamical heating
due to Galactic tides.}
\end{figure}

\section{Conclusions}

We have taken a large homogeneous sample of published velocity dispersion profiles for 40 Galactic globular clusters, to study
the first order characteristics of the kinematics of these systems in their outer low acceleration regions.

We find that a simple central Gaussian region and a flat asymptote, $\sigma_{\infty}$, are sufficient to very accurately model
all the observed velocity dispersion profiles treated.

Estimating masses from stellar population models not including any dynamical assumptions, then allows to investigate the empirical
relation existing between $\sigma_{\infty}$ and $M$. While no {\it a priori} relation between these two parameters was assumed or
forced, the results show a very clear signal consistent with the isothermal pressure supported equivalent of the baryonic Tully-Fisher
relation in disk galaxies, of  $\sigma_{\infty} (km s^{-1})=0.2(M/M_{\odot})^{0.25}$.

This result becomes a further constraint towards understanding the formation and evolution of globular clusters, under either
a Newtonian framework, or within any alternative theory of gravity.

\begin{table}
\begin{flushleft}
  \caption{Parameters for the globular clusters treated.}
  \begin{tabular}{l@{\:}|r@{\:}|r@{\:}|r@{\:}|c@{\:}|c@{\:}|c@{\:}} 
  \hline
  \hline

GC &$\sigma_{1} (km/s)$&$\sigma_{\infty} (km/s)$ &$R_{\sigma}$ (pc)& $log(M /M_{\odot})$ \\
  \hline

NGC 104&5.49$\pm$0.35&4.22$\pm$0.33&13.87$\pm$1.12&6.06$^{+0.06}_{-0.07}$ \\
NGC 362&3.96$\pm$0.28&2.87$\pm$0.30&10.68$\pm$1.12&5.61$^{+0.07}_{-0.08}$ \\
NGC 1261&2.03$\pm$0.38&1.92$\pm$0.40&15.41$\pm$3.51&5.33$^{+0.07}_{-0.07}$ \\
NGC 1851&3.06$\pm$0.44&3.22$\pm$0.29&10.63$\pm$1.95&5.56$^{+0.07}_{-0.07}$ \\
NGC 1904&1.88$\pm$0.19&2.21$\pm$0.17&12.54$\pm$1.74&5.33$^{+0.06}_{-0.06}$ \\
NGC 2808&5.32$\pm$0.83&4.75$\pm$0.93&14.91$\pm$3.12&5.99$^{+0.07}_{-0.08}$ \\
NGC 3201&1.31$\pm$0.26&2.62$\pm$0.27&6.93$\pm$1.70&5.17$^{+0.06}_{-0.06}$ \\
NGC 4372&1.96$\pm$0.24&2.58$\pm$0.27&9.72$\pm$1.63&5.29$^{+0.06}_{-0.06}$ \\
NGC 4590&1.74$\pm$0.09&1.89$\pm$0.06&9.09$\pm$0.64&5.12$^{+0.06}_{-0.06}$ \\
NGC 4833&1.25$\pm$0.55&3.48$\pm$0.27&3.98$\pm$1.97&5.45$^{+0.06}_{-0.06}$ \\
NGC 5139&8.92$\pm$0.58&6.90$\pm$0.38&14.87$\pm$1.36&6.29$^{+0.06}_{-0.06}$ \\
NGC 5272&4.26$\pm$0.40&2.48$\pm$0.28&12.78$\pm$1.69&5.74$^{+0.06}_{-0.06}$ \\
NGC 5904&3.75$\pm$0.23&2.83$\pm$0.24&13.94$\pm$1.34&5.75$^{+0.07}_{-0.07}$ \\
NGC 5927&2.35$\pm$0.95&3.32$\pm$1.04&12.79$\pm$5.51&5.50$^{+0.07}_{-0.07}$ \\
NGC 5986&3.49$\pm$1.11&4.43$\pm$0.44&4.64$\pm$1.51&5.56$^{+0.06}_{-0.06}$ \\
NGC 6093&4.02$\pm$0.83&5.39$\pm$0.42&2.58$\pm$0.64&5.47$^{+0.06}_{-0.06}$ \\
NGC 6121&2.22$\pm$0.61&2.27$\pm$0.65&5.16$\pm$1.29&5.11$^{+0.07}_{-0.08}$ \\
NGC 6171&2.13$\pm$0.38&1.84$\pm$0.42&7.67$\pm$1.83&5.10$^{+0.06}_{-0.08}$ \\
NGC 6205&3.38$\pm$0.34&3.57$\pm$0.34&12.14$\pm$1.71&5.61$^{+0.06}_{-0.06}$ \\
NGC 6218&1.61$\pm$0.23&2.55$\pm$0.19&4.32$\pm$0.86&5.12$^{+0.07}_{-0.06}$ \\
NGC 6273&7.09$\pm$0.42&4.73$\pm$0.32&6.82$\pm$0.56&5.83$^{+0.06}_{-0.06}$ \\
NGC 6304&2.15$\pm$0.67&3.26$\pm$0.84&3.25$\pm$1.56&5.24$^{+0.06}_{-0.07}$ \\
NGC 6341&3.85$\pm$0.65&2.33$\pm$0.50&12.63$\pm$2.44&5.46$^{+0.06}_{-0.07}$ \\
NGC 6388&6.02$\pm$0.54&5.50$\pm$0.62&11.34$\pm$1.74&6.08$^{+0.06}_{-0.07}$ \\
NGC 6397&1.62$\pm$0.11&2.96$\pm$0.12&3.94$\pm$0.40&4.83$^{+0.06}_{-0.06}$ \\
NGC 6402&5.34$\pm$1.69&3.99$\pm$2.04&11.15$\pm$4.79&5.85$^{+0.07}_{-0.07}$ \\
NGC 6441&7.21$\pm$0.87&8.98$\pm$0.31&4.69$\pm$0.49&6.19$^{+0.06}_{-0.07}$ \\
NGC 6517&3.56$\pm$1.69&6.56$\pm$0.32&2.68$\pm$1.29&5.51$^{+0.07}_{-0.07}$ \\
NGC 6541&2.49$\pm$0.42&3.39$\pm$0.47&8.83$\pm$1.80&5.59$^{+0.06}_{-0.06}$ \\
NGC 6544&2.61$\pm$0.38&3.55$\pm$0.41&2.65$\pm$0.40&4.97$^{+0.06}_{-0.06}$ \\
NGC 6553&2.48$\pm$1.20&6.30$\pm$0.16&1.67$\pm$0.51&5.54$^{+0.07}_{-0.08}$ \\
NGC 6569&1.62$\pm$0.14&4.52$\pm$0.07&4.07$\pm$0.40&5.59$^{+0.06}_{-0.07}$ \\
NGC 6624&2.75$\pm$0.50&3.52$\pm$0.24&1.76$\pm$0.39&5.36$^{+0.07}_{-0.07}$ \\
NGC 6626&4.74$\pm$0.43&3.81$\pm$0.49&5.70$\pm$0.72&5.47$^{+0.07}_{-0.06}$ \\
NGC 6656&3.62$\pm$0.18&4.29$\pm$0.19&6.48$\pm$0.50&5.58$^{+0.06}_{-0.06}$ \\
NGC 6723&2.89$\pm$0.81&2.23$\pm$0.87&8.43$\pm$2.85&5.37$^{+0.07}_{-0.08}$ \\
NGC 6752&2.99$\pm$0.27&3.47$\pm$0.17&5.55$\pm$0.58&5.28$^{+0.06}_{-0.06}$ \\
NGC 6838&1.48$\pm$0.87&2.57$\pm$0.11&0.89$\pm$0.35&4.54$^{+0.06}_{-0.07}$ \\
NGC 7078&6.47$\pm$0.56&3.09$\pm$0.31&8.15$\pm$0.94&5.85$^{+0.06}_{-0.07}$ \\
NGC 7099&2.10$\pm$0.44&2.12$\pm$0.40&8.35$\pm$2.80&5.15$^{+0.06}_{-0.06}$ \\

 \hline
\end{tabular} 
After the globular cluster identification column, the following three entries give the parameters of the fits to equation(1) for
the observed projected velocity dispersion profiles of Baumgard et al. (2019) and their confidence intervals. The last column gives
total stellar mass estimates from the stellar population modelling of McLaughlin \& van der Marel (2005).

\end{flushleft}
\end{table}

\section*{acknowledgements}
The authors acknowledge the constructive criticism of an anonymous referee as important towards
reaching a more complete and clear final version. Xavier Hernandez acknowledges financial assistance
from UNAM DGAPA grant IN104517 and CONACYT.

\end{document}